\documentclass[%
 reprint,
superscriptaddress,
showpacs,
showkeys,
 amsmath,amssymb,
 aps,
pre,
]{revtex4-1}

\usepackage{graphicx}
\usepackage{dcolumn}
\usepackage{bm}
\usepackage{hyperref}

\usepackage{float}
\usepackage{braket}
\usepackage{epstopdf}
\usepackage{pgfplots}

\newcommand{\Curl}{\bm{\nabla}\times}
\newcommand{\Div}{\bm{\nabla}\cdot}
\newcommand{\ii}{\mathrm{i}}
\newcommand{\qperp}{\bm{q}_\perp}
\newcommand{\ee}{\mathrm{e}}
\newcommand{\hatq}{\bm{\hat{q}}}
\newcommand{\hatx}{\bm{\hat{x}}}
\newcommand{\haty}{\bm{\hat{y}}}
\newcommand{\hatz}{\bm{\hat{z}}}
\newcommand{\hatqp}{\bm{\hat{q}}_\perp}
\newcommand{\hats}{\bm{\hat{s}}}
\newcommand{\hatp}{\bm{\hat{p}}}

\newcommand{\IM}{\mathrm{Im}}
\newcommand{\RE}{\mathrm{Re}}

\newcommand{\expp}{\ee^{\ii2\gamma_2d}}

\newcommand{\Dp}{1-\mr_{21}^p\mr_{23}^p\ee^{\ii2\gamma_2 d}}
\newcommand{\Ds}{1-\mr_{21}^s\mr_{23}^s\ee^{\ii2\gamma_2 d}}

\newcommand{\mt}{T}
\newcommand{\mr}{R}

\newcommand{\denompo}{\gamma_1\epsilon_2+\gamma_2\epsilon_1+\gamma_1\gamma_2\frac{\sigma_0}{\varepsilon_0\omega}}%
\newcommand{\denompd}{\gamma_2\epsilon_3+\gamma_3\epsilon_2+\gamma_2\gamma_3\frac{\sigma_D}{\varepsilon_0\omega}}
\newcommand{\denomso}{\gamma_1+\gamma_2+\mu_0\omega\sigma_0}
\newcommand{\denomsd}{\gamma_2+\gamma_3+\mu_0\omega\sigma_D}

\begin{document}


\title{Radiative heat transfer as a Landauer-B\"{u}ttiker problem}

\author{Han Hoe \surname{Yap}}
\affiliation{NUS Graduate School for Integrative Sciences and Engineering, Singapore 117597, Republic of Singapore.}
\author{Jian-Sheng \surname{Wang}}%
\affiliation{%
Department of Physics, National University of Singapore, Singapore 117551, Republic of Singapore.
}%

\date{16 September 2016}

\begin{abstract}
We study the radiative heat transfer between two semi-infinite half-spaces, bounded by conductive surfaces in contact with vacuum. This setup is interpreted as a four-terminal mesoscopic transport problem. The slabs and interfaces are viewed as bosonic reservoirs, coupled perfectly to a scattering center consisting of the two interfaces and vacuum. Using Rytov's fluctuational electrodynamics and assuming Kirchhoff's circuital law, we calculate the heat flow in each bath. This allows for explicit evaluation of a conductance matrix, from which one readily verifies B\"{u}ttiker symmetry. Thus, radiative heat transfer in layered media with conductive interfaces becomes a Landauer-B\"{u}ttiker transport problem.
\end{abstract}

\pacs{44.40.+a, 73.23.Ad}
\keywords{Thermal radiation, Ballistic transport}
\maketitle


\section{\label{sec:intro}INTRODUCTION}
\begin{figure}[H]
	\centering
	\includegraphics[scale=1]{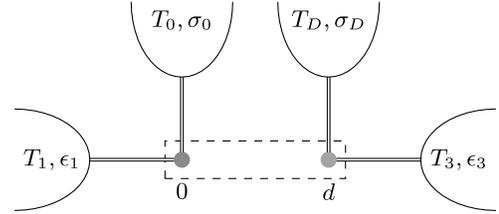}
	\caption{Layered media 1, 3 with conductive interfaces 0, $D$ as a four-terminal Landauer-B\"{u}ttiker transmission problem. Planar symmetry reduces the interfaces to two dots, serving as a scattering center enclosed by the dashed rectangle.}\label{fig:LB}
\end{figure}

The study of thermal radiation began in late nineteenth and early twentieth century. In 1906, Planck evoked in his magisterial work \cite{Planck} the notion of far-field assumption, namely the wavelengths of interest are much smaller compared to the spatial extensions of the objects considered. Combined with Kirchhoff's black-body idealization, the radiative heat transfer between two bodies in the far-field regime is then bounded by Stefan-Boltzmann law. In 1969, Hargreaves \cite{Hargreaves} reported an anomalous radiative heat transfer between two chromium plates separated at a distance less than 3 $\mu$m, surpassing the black-body limit. Since then, there has been a surge in research activities \cite{Volokitin2,Rousseau,Kralik,Song,Song2,Chen} concerning radiative heat transfer in the near-field regime, from Polder and van Hove's pioneering work \cite{PvH} in 1971, to the experimental measurements of St-Gelais et al. \cite{StGelais} in 2016. Of particular interest is the geometry of planar layered media \cite{Francoeur,Greffet}, where the case of metamaterial \cite{meta1,meta2} and coated media \cite{coat1,coat2} were further investigated. \paragraph*{}Meanwhile, recent years have seen growing interests in interdisciplinary studies \cite{transistor,transistor2,QHE,meso} involving both thermal radiation and mesoscopic transport, which is not surprising considering their similarities. In this work, we wish to strengthen this connection by revisiting the case of two layered media separated by vacuum with conductive interfaces. While earlier works \cite{coat1,coat2} were targeted at the tunability of radiative transfer when bulk dielectrics are covered by graphene sheets, here we focus on the energy-balance aspect. In so doing, we find it natural---at least for our system of interest---to regard radiative heat transfer as a four-terminal transport problem. Once a set of detailed-balance conditions are checked, radiative heat transfer becomes a Landauer-B\"{u}ttiker model, as shown in Fig.~\ref{fig:LB}. 
\paragraph*{}The summary of this paper is as follows. We first fix notations and describe the model. Electromagnetic propagation in layered media with conductive interfaces is then discussed. Next, we outline the steps to calculate the radiative heat transfer. The main novelty of this work starts from Sec.~\ref{subsec:Landauer}, where we present a four-terminal Landauer-B\"{u}ttiker viewpoint for the problem of interest.
The electromagnetic energy in each bath is related to the Bose function via a conductance matrix. This allows for more transparent energy balance in each terminal (bulk medium or interface), elucidates the ballistic aspect of fluctuational electrodynamics, provides another justification for neglecting propagating mode in near-field radiative transfer. We then illustrate with an example of a graphene-coated gold radiating to a suspended graphene sheet. 

\section{\label{sec:mainbody}MODEL AND SOLUTION}
We work in the framework of fluctuational electrodynamics \cite{Rytov}, where sources are considered as thermal noises, whose auto-correlation functions can be deduced from the fluctuation-dissipation theorem. We consider layered media aligned along $z$ direction, whose cross-sections are modeled to be infinitely large. We are interested in steady-state transport, so all time arguments will be Fourier-transformed to frequency. We adopt the standard convention of Fourier transform which results in the following replacement: $\frac{\partial}{\partial t}\to -\ii\omega$ and $\bm{\nabla}\to \ii\bm{q}$, where $\bm{q}$ is a three-dimensional wave-vector. A point in space will be denoted by $\bm{r}=(x,y,z)$ and in plane $\bm{R}=(x,y)$. The perpendicular subscript $\perp$ indicates transverse to $z$ direction. We use $\varepsilon_0$ to denote the permittivity of free space and $\epsilon_i$ to denote the dimensionless dielectric constant of medium $i$.
\subsection{\label{subsec:model}Model}
We consider two semi-infinite half-spaces, kept at temperatures $T_{1,3}$ with local dielectric functions $\epsilon_{1,3}(\bm{r},\omega)=\epsilon_{1,3}(\omega)$, separated by vacuum at a distance $d$ away from each other. The interfaces are equipped with additional properties: they are given temperatures $T_{0,D}$ and local conductivities $\sigma_{0,D}(\bm{R},\omega)=\sigma_{0,D}(\omega)$.
\begin{figure}[H]
	\centering
	\includegraphics[scale=1]{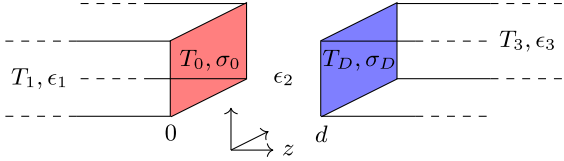}
	\caption{Two semi-infinite slabs with temperatures $T_{1,3}$ and dielectric functions $\epsilon_{1,3}$ separated by a vacuum gap ($\epsilon_2=1$) at a distance $d$ apart. The interfaces in contact with vacuum are at temperatures $T_{0,D}$ with conductivities $\sigma_{0,D}$. The planar cross-sections should be understood to be infinitely large.}
\end{figure}
Ordinarily, the interfaces---being in contact with the bulk---are naturally given the same temperatures as their respective slabs. A Landauer-B\"{u}ttiker perspective shows that these temperatures can in principle be different. Finally, we consider non-magnetic linear isotropic media, so that the constitutive relations are given by $\bm{H}=\bm{B}/\mu_0$, $\bm{D}=\varepsilon_0\epsilon\bm{E}$.
\subsection{\label{subsec:Maxwell}Maxwell Equations}For electromagnetic propagation in media, the inhomogeneities in Maxwell equations are excessive charges and currents, which in the present context are assumed to originate from thermal fluctuations. In what follows we denote by $\rho(\bm{r},\omega)$ ($\Sigma_{0,D}(\bm{R},\omega)$) the volume (surface) charge density, and $\bm{J}(\bm{r},\omega)$ ($\bm{K}_{0,D}(\bm{R},\omega)$) the volume (surface) current density. We need to solve the following Maxwell equations:
\begin{align}
\Div\bm{D} &=\rho+ \Sigma_0\delta(z)+\Sigma_D\delta(z-d), \label{eq:Gauss} \\
\Div \bm{B} &= 0, \label{eq:Gilbert} \\
\Curl \bm{H}+ \ii\omega\bm{D} &= \bm{J} + \bm{K}_0\delta(z)+\bm{K}_D\delta(z-d), \label{eq:Ampere} \\
\Curl \bm{E} - \ii\omega\bm{B} &=0. \label{eq:Faraday}
\end{align}
In the above, not all equations are independent. In particular, Gauss' law \eqref{eq:Gauss} and Amp\`{e}re's law \eqref{eq:Ampere} are related by the continuity equations $\ii\omega\rho = \Div \bm{J}$ and $\ii\omega \Sigma_{0,D} = \bm{\nabla}_\perp\cdot \bm{K}_{0,D}$, where $\bm{\nabla}_\perp= \hatx\frac{\partial}{\partial x} + \haty\frac{\partial}{\partial y}$ is a ``transverse divergence''.
\subsubsection{Single interface}
Let us begin by considering one half-space with dielectric constant $\epsilon_1$ in contact with vacuum at $z=0$. The dielectric function then reads: $\epsilon(z,\omega)=\theta(z)\epsilon_2+\theta(-z)\epsilon_1(\omega)$, where for more symmetric expressions we denoted by $\epsilon_2$ the dielectric constant of vacuum (=1). Such a form, together with the presence of delta functions in \eqref{eq:Gauss} and \eqref{eq:Ampere}, suggest the following form of solution: $\bm{E}(\bm{r})=\theta(z)\bm{E}_>(\bm{R},z)+\delta(z)\bm{E}_{\delta}(\bm{R})+\theta(-z)\bm{E}_<(\bm{R},z)$. The linear independence of $\theta(\pm z)$ and $\delta(z)$ then leads to $\bm{E}_{\delta}=0$, i.e. the electric field is at worst discontinuous. Similar reasoning follows for the magnetic induction $\bm{B}$. One then arrives at the following saltus conditions \cite{Sipe}:
\begin{align}
	\hatz\cdot(\epsilon_2\bm{E}_>-\epsilon_1\bm{E}_<) &= \frac{\Sigma_0}{\varepsilon_0}, \\
	\hatz\cdot(\bm{B}_>-\bm{B}_<) &= 0, \\
	\hatz\times(\bm{B}_>-\bm{B}_<)&= \mu_0 \bm{K}_0, \label{eq:salt3} \\
	\hatz\times(\bm{E}_>-\bm{E}_<) &= 0, \label{eq:salt4}
\end{align}
as well as two sets of Maxwell equations: a homogeneous one in vacuum, and another in $z<0$ bulk with thermal fluctuation $\rho,\bm{J}$:
\begin{align}
	\Div (\epsilon_1\bm{E}_<) &= \frac{\rho}{\varepsilon_0}, \label{eq:bulk1} \\
	\Div\bm{B}_< &= 0, \\
	\Curl \bm{B}_<+\frac{\ii\omega}{c^2}\epsilon_1 \bm{E}_< &= \mu_0\bm{J}, \\
	\Curl \bm{E}_< -\ii\omega\bm{B}_< &= 0.	\label{eq:bulk4}
\end{align}
We briefly outline one way \cite{Greffet} of solving Eqs. \eqref{eq:bulk1}---\eqref{eq:bulk4}. Dropping the subscripts, one begins with the potentials $\bm{A},\phi$ and works in Lorenz gauge $\Div\bm{A}={\ii\omega}\epsilon\phi/{c^2}$, so that the vector potential satisfies an inhomogeneous Helmholtz equation: $\left[ \bm{\nabla}^2+{\omega^2}\epsilon/{c^2}\right]\bm{A}=-\mu_0\bm{J}$ and the electric field reads: $\bm{E}=\ii\omega\left[1+k^{-2}{\bm{\nabla}\bm{\nabla}}\right]\bm{A}$, with $k^2={\omega^2}\epsilon/{c^2}$. Denoting the wave-vector by:
\begin{equation} 
\bm{q}=q_x \hatx+q_y\haty+\gamma\hatz:= \qperp + \gamma\hatz, \label{eq:wavevector}
\end{equation}
the partial Fourier-transformed electric field is then given by:
\begin{equation}  \bm{E}(\qperp,z)=\ii\omega\mu_0\int_{-\infty}^{z}\mathrm{d}z'\;\bm{G}^E(\qperp,z-z')\cdot\bm{J}(z'),\label{eq:E}
\end{equation}where the electric Green's dyadic is \cite{Eskin}:
\begin{equation}
\begin{split}  \label{eq:GE}
\bm{G}^E&(\qperp,z)=-\frac{1}{k^2}\delta(z)\hatz\hatz\\&+\frac{\ii}{2\gamma}\left[\theta(z)(\bm{1}-\hatq_+\hatq_+)\ee^{\ii\gamma z}+\theta(-z)(\bm{1}-\hatq_-\hatq_-)\ee^{-\ii\gamma z}\right].
\end{split} 
\end{equation}
In Eq. \eqref{eq:GE}, the $z$ component of wave-vector, $\gamma$, is constrained by $\gamma=\sqrt{{\omega^2}\epsilon/{c^2}-\qperp^2}$, square root taken such that $\IM(\gamma)>0$ to guarantee bounded electric fields. Also, we defined the unit vectors $\hatq_\pm=(\qperp\pm\gamma\hatz)/q$, which can be interpreted as forward or backward moving wave-vector. They are unit in the sense that $\hatq_\pm\cdot\hatq_\pm=1$. As for the delta function, unlike the previous two terms, it is not accompanied by exponential $\ee^{\pm\ii\gamma z}$, representing thus a static electric field. Alternatively, if we compute the magnetic induction corresponding to this delta function, the Poynting vector is $\qperp$ independent, but points along $\hatqp$, so it integrates to zero and does not contribute in heat transfer.
\subsubsection{$s$ and $p$ polarizations}
Electromagnetic propagation in layered media singles out $z$ axis as a special direction $\hatz$. Granted planar symmetry, for each Fourier component we have another vector, $\hatqp={\qperp}/{|\qperp|}$. They can be completed by a third vector $\hats(\qperp)$ to give a right-hand triple $(\hats,\hatqp,\hatz)$. By construction this $\hats$ is orthogonal to the total wave-vector $\hatq$ given by \eqref{eq:wavevector}. The solution of electric field, \eqref{eq:E}---\eqref{eq:GE}, implies a posteriori that for each Fourier component $\qperp$, the electric field is transverse: $\hatq_\pm\cdot\bm{E}=0$. Thus with the unit-vector $\hats$ we define $\hatp_\pm:=\hats\times\hatq_\pm$, obtaining another right-hand triple $(\hats,\hatq_\pm,\hatp_\pm)$. Notice that the $p$ polarization $\hatp$ and wave-vector $\hatq$ are generally complex vectors. Therefore electric fields are described by $\bm{E}_\pm=E_{s_\pm}\hats + E_{p_\pm}\hatp_\pm$, and from Faraday's law \eqref{eq:Faraday} we can express the magnetic field in terms of the amplitudes, i.e. $\bm{H}_\pm={\sqrt{\epsilon}} (E_{p_\pm}\hats - E_{s_{\pm}}\hatp_\pm)/({\mu_0 c})$.
\subsection{\label{subsec:conductive}Single Conductive Interface}
\begin{figure}[H]
	\centering
	\includegraphics[scale=1]{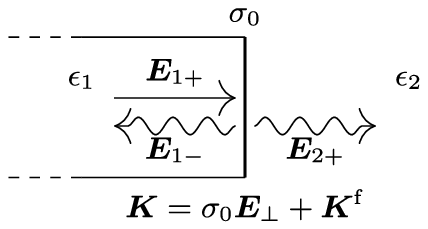}
	\caption{Scattering-emission problem: given incident electric field $\bm{E}_{1+}$, fluctuating surface current $\bm{K}^\mathrm{f}$, demanding current response $\sigma_0\bm{E}_\perp$, find the outgoing fields $\bm{E}_{1-}$ and $\bm{E}_{2+}$.}\label{fig:charged}
\end{figure}

We now turn to the role played by conductive interfaces. Again we first focus on one single interface and derive its consequences. Dropping the location subscripts, we write the surface current, residing purely on the interfaces, as: $\bm{K}=\bm{K}^\sigma+\bm{K}^\mathrm{f}$, where $\bm{K}^\sigma$ is a deterministic response to electric fields, $\bm{K}^\mathrm{f}$ is the fluctuating part that arises from thermal motions. We wish to solve the following problem, as shown in Fig.~\ref{fig:charged}.
Saltus conditions \eqref{eq:salt3}--\eqref{eq:salt4} then assert:
\begin{align}
\begin{split}
&\hats \left[-\frac{\gamma_2}{\omega}E_{s_{2+}}+\frac{\gamma_1}{\omega}(E_{s_{1+}}-E_{s_{1-}})\right] \\
+&\hatqp \left[\frac{\sqrt{\epsilon_2}}{c}E_{p_{2+}}-\frac{\sqrt{\epsilon_1}}{c}(E_{p_{1+}}+E_{p_{1-}})\right]=\mu_0 \bm{K},
\end{split}\label{eq:salt3b}
\end{align}
\begin{align}
\begin{split}
&\hats \left[E_{s_{2+}}-(E_{s_{1+}}+E_{s_{1-}})\right] \\
+&\hatqp \left[\frac{\gamma_2}{q_2}E_{p_{2+}}-\frac{\gamma_1}{q_1}(E_{p_{1+}}-E_{p_{1-}})\right] = 0.\label{eq:salt4b}
\end{split} 
\end{align}
It is worthwhile to mention that the conductive interfaces we consider are truly two-dimensional, so they are not modeled by thin films (rectangular solid of small but non-zero thickness).
\subsubsection{Modified Fresnel law}
We first consider the scattering aspect of Fig.~\ref{fig:charged} by putting aside the fluctuating part. The electric-field-induced current is given by a planar-isotropic phenomenological Ohm's law: $\bm{K}^\sigma=\sigma_0\bm{E}_\perp$. This expression is meaningful because Eq. \eqref{eq:salt4} implies the continuity of the in-plane component of electric field (denoted by $\bm{E}_\perp$). However, the $p$ polarized electric field generally has non-zero $z$ component. Hence, for each Fourier component $\qperp$ we write $\bm{K}^\sigma=K^\sigma_s\hats + K^\sigma_{q_\perp}\hatqp$, so that the current responses due to the electric field read $K^\sigma_s = \sigma_0 E_{s_{2+}}$ and $K^\sigma_{q_\perp}=\sigma_0 \gamma_2E_{p_{2+}}/q_2$. From the saltus conditions \eqref{eq:salt3b}---\eqref{eq:salt4b}, we are led to the following modified Fresnel coefficients \cite{Volokitin}:
\begin{align}
&\mt^s_{12}=\frac{2\gamma_1}{\gamma_1+\gamma_2+\mu_0\omega\sigma_0},\label{eq:mFresnels}\\ &\mr^s_{12} = \frac{\gamma_1-\gamma_2-\mu_0\omega\sigma_0}{\gamma_1+\gamma_2+\mu_0\omega\sigma_0}, \\
&\mt^p_{12}=\frac{2\gamma_1\sqrt{\varepsilon_1}\sqrt{\varepsilon_2}}{\gamma_1\varepsilon_2+\gamma_2\varepsilon_1+\gamma_1\gamma_2\frac{\sigma_0}{\varepsilon_0\omega}}, \\& \mr_{12}^p=\frac{\gamma_1\varepsilon_2-\gamma_2\varepsilon_1+\gamma_1\gamma_2\frac{\sigma_0}{\varepsilon_0\omega}}{\gamma_1\varepsilon_2+\gamma_2\varepsilon_1+\gamma_1\gamma_2\frac{\sigma_0}{\varepsilon_0\omega}}, \label{eq:mFresnelp}
\end{align}
so that as a function of the incoming electric field $\bm{E}_{1+}$, the transmitted and reflected fields are respectively given by $E_{s/p2+}=\mt^{s/p}_{12}E_{s/p1+}$ and $E_{s/p1-}=\mr^{s/p}_{12}E_{s/p1+}$. When $\sigma_0=0$, we recover the usual Fresnel law for dielectric media \cite{Jackson}. 
\subsubsection{Fluctuating surface current}
We now incorporate the thermal fluctuation $\bm{K}^\mathrm{f}$ in the surface current. Here, we are interested in the emission aspect of Fig.~\ref{fig:charged}, so in Eqs. \eqref{eq:salt3b}---\eqref{eq:salt4b} we put $\bm{E}_{1+}=0$ but still include the response part $\bm{K}^\sigma$. This yields the following emitted electric fields:
\begin{align}
E^\mathrm{f}_{s_{1-}}&=-\frac{\mt^s_{21}}{2\gamma_2}\mu_0\omega K^\mathrm{f}_s, \\
E^\mathrm{f}_{p_{1-}}&=-\frac{\mt^p_{21}}{2\gamma_2}\mu_0\omega\left[\frac{\gamma_2}{q_2}K^\mathrm{f}_{q_\perp}\right], \\
E^\mathrm{f}_{s_{2+}} &=-\frac{\mt^s_{12}}{2\gamma_1}\mu_0\omega K^\mathrm{f}_s, \\
E^\mathrm{f}_{p_{2+}}&=-\frac{\mt^p_{12}}{2\gamma_1}\mu_0\omega\left[-\frac{\gamma_1}{q_1}K^\mathrm{f}_{q_\perp}\right].
\end{align}
Thanks to the linearity of Maxwell equations, the outgoing electric fields are simply the sum of scattering and emission problem treated separately: $E_{s/p2+}=\mt^{s/p}_{12}E_{s/p1+}+E_{s/p1+}^\mathrm{f}$ and $E_{s/p1-}=\mr^{s/p}_{12}E_{s/p1+}+E_{s/p1-}^\mathrm{f}$.
\subsection{Two Conductive Interfaces}
We thus arrive at the conclusion that conductive interface generates its own field and modifies Fresnel law. Returning to the problem of interest, we need to solve a scattering problem with two half-spaces and two interfaces. We follow Sipe's approach \cite{Sipe}: First, write the forward and backward moving fields in medium 1 as a vector, say $v_1$. Then, reorganize the saltus conditions to obtain a single-interface transfer matrix, say $M_{1\to2}$ for crossing from medium 1 to 2. Next, propagation in the medium 2 across a distance $d$ can be described by a simple matrix $P_2(d)$. Finally, the fields on the third region are obtained by acting the matrices consecutively on: $v_3=M_{2\to3}P_2(d)M_{1\to2}v_1$. In this approach, so long as for one interface we have Fresnel law, it does not matter whether it is the modified coefficients \eqref{eq:mFresnels}---\eqref{eq:mFresnelp} or the usual ones (by setting $\sigma_0=0$).

\subsection{\label{subsec:rht}Radiative Heat Transfer}
In this section we consider the energy transport by electromagnetic waves. For a clearer analogy to mesoscopic transport, we express the autocorrelation functions in Fourier domain $\qperp$. This is possible because there is no spatial dispersion in the dissipative properties $\epsilon,\sigma$. The first task is to obtain an expression for electromagnetic energy flux density (Poynting vector) from fields expressed in frequency. Then, we need to find the autocorrelations of the fluctuating currents.
\subsubsection{Poynting vector}
We consider coherent emission, i.e. random fields $F$ in frequency domain are delta-correlated: $\braket{F(\omega)F(\omega')}\propto \delta(\omega+\omega')$. This is equivalent to saying that $F$ is stationary in the wide sense \cite{Mandel}: $\braket{F(t)F(t')}=\braket{F(t-t')F(0)}$. Then, the expression of Poynting vector is given by:
\begin{equation}
\braket{\bm{S}(\bm{r})}=\int_0^{\infty}\frac{\mathrm{d}\omega}{2\pi}\bm{S}(\bm{r},\omega),\label{eq:Poynting}
\end{equation}
where $\bm{S}(\bm{r},\omega)=2\braket{\RE[\bm{E}(\bm{r},\omega)\times\bm{H}^*(\bm{r},\omega)]}$ can be seen as a spectral Poynting vector, and the $\braket{\dots}$ is an average over realizations of the random currents. There is no time argument in Eq. \eqref{eq:Poynting}, which is natural because we consider steady-state transport. Thanks to planar symmetry, only the $z$ component contributes and the solution depends at most on $z$ but not $\bm{R}$.

\subsubsection{\label{subsubsec:fdt}Fluctuation-dissipation theorem}
We model the thermal fluctuations by random currents $\bm{J},\bm{K}$, so their statistical properties must be related to the temperature. This relation is given by the fluctuation-dissipation theorem \cite{Landau,Eckhardt,Kubo}:
\begin{equation}
\begin{split}
&\braket{J_i(\bm{r},\omega)J_j^*(\bm{r}',\omega')}=\delta(\omega-\omega')\delta(\bm{r}-\bm{r}')\delta_{ij} \\
&\times \hbar\omega^2\varepsilon_0\IM[\epsilon(\omega)]\coth\left[\frac{\beta\hbar\omega}{2}\right]. \label{eq:JJ}
\end{split}
\end{equation}
For the interfaces, in the prescriptions of fluctuation-dissipation theorem, we simply replace the bulk current by its surface counterpart $\bm{K}$, and use conductivity $\sigma$ in place of dielectric constant $\epsilon$. We obtain:
\begin{equation}
\begin{split} 
&\braket{K_i(\bm{R},\omega)K_j^*(\bm{R}',\omega')}=\delta(\omega-\omega')\delta(\bm{R}-\bm{R}')\delta_{ij} \\
&\times \hbar\omega\RE[\sigma(\omega)]\coth\left[\frac{\beta\hbar\omega}{2}\right], \label{eq:KK}
\end{split}
\end{equation}
where $i,j$ are any two in-plane components.

\section{\label{subsec:Landauer}Landauer-B\"{u}ttiker Formalism}
\subsection{Overview}
An important principle in mesoscopic transport reads ``transmission is scattering'' \cite{Datta}. Similarly, when studying radiative transfer in layered media, one solves an electromagnetic wave scattering problem before calculating the energy transmitted. In mesoscale, electrons are described by wavefunctions $\psi(\bm{x})$, whose amplitude-squared $|\psi(\bm{x})|^2$ represents a quantity that can be observed. Analogously, here the electromagnetic field alone is fluctuating. Thus, it averages to zero, but electromagnetic energy---being quadratic in the noise $\bm{J},\bm{K}$---is in general not zero.
\paragraph*{}To consider non-equilibrium transport, one needs baths: a large system capable of supplying particles or energy without being appreciably affected. In our problem, the obvious candidates for baths are the two bulk media. Adding two conductive interfaces (which radiate and modify the scattering) suggests the scenario as depicted in Fig.~\ref{fig:LB}, where the surfaces now serve both as baths (since they radiate) and scatterers (since they are the reasons of electromagnetic wave scattering). By imposing Kirchhoff's circuital law, i.e. energy conservation at nodes 0 and $D$, we show that the radiative heat transfer in layered media with conductive interfaces is essentially a four-terminal Landauer-B\"{u}ttiker transmission problem. With the convention that energy entering the bath is positive, we write:
\begin{equation}
\begin{bmatrix}
S_1 \\ S_0 \\ S_D \\ S_3
\end{bmatrix}=\hbar\omega
\begin{bmatrix}
G_{11} & G_{01} & G_{D1} & G_{31} \\
G_{10} & G_{00} & G_{D0} & G_{30} \\
G_{1D} & G_{0D} & G_{DD} & G_{3D} \\
G_{13} & G_{03} & G_{D3} & G_{33}
\end{bmatrix}
\begin{bmatrix}
N_1 \\ N_0 \\ N_D \\ N_3
\end{bmatrix}.\label{eq:mat}
\end{equation}
In above, $S_\alpha(\qperp,\omega)$ is the Fourier-resolved energy in link $\alpha$. $G_{\alpha\delta}$ is a dimensionless quantity known as conductance (between terminals $\alpha,\delta$) or spectral function. We stress that these variables depend on polarization, whose indices are omitted for simpler notation. $N_\alpha(\omega)=(\ee^{\beta_\alpha\hbar\omega}-1)^{-1}$ is the Bose function of bath $\alpha$. By the imposed energy balance, we have $G_{\alpha\alpha}=-\sum_{\delta\neq\alpha} G_{\delta\alpha}$, as well as B\"{u}ttiker symmetry $G_{\alpha\delta}=G_{\delta\alpha}$ that can be checked once the matrix elements are evaluated explicitly (see Appendix~\ref{app:Bulk}---\ref{app:Interface}). To calculate the total energy $\mathcal{S}_\alpha$ in link $\alpha$, one would then integrate over channels $\qperp$ and frequencies $\omega$:
\begin{equation}
\begin{split} 
\mathcal{S}_\alpha = \sum_{\delta\neq\alpha}\int_0^{\infty}&\frac{\mathrm{d}\omega}{2\pi}\hbar\omega\left[N_\delta(\omega)-N_\alpha(\omega)\right] \\ &\times \sum_{\lambda\in\{s,p\}} \int_{\mathbb{R}^2}\frac{\mathrm{d}^2\qperp}{(2\pi)^2}G^\lambda_{\alpha\delta}(\qperp,\omega), \label{eq:rht}
\end{split} 
\end{equation}
where we restored the polarization index $\lambda$ on the conductance $G_{\alpha\delta}^\lambda$.

\subsection{Discussion}
\subsubsection{Detailed balance}
Hidden under B\"{u}ttiker symmetry is a set of detailed-balance conditions satisfied by the products of the (modulus-squared) electric field and the dissipative term. Roughly speaking, one has $\Pi_{\alpha}I_{\alpha\to\delta}=\Pi_{\delta}I_{\delta\to\alpha}$, where $I_{\alpha\to\delta}$ is the modulus-squared electric field in link $\delta$ due to fluctuation in bath $\alpha$, and $\Pi_{\alpha}$ is the dissipation in bath $\alpha$. We refer the reader to Appendix \ref{app:Bulk}---\ref{app:Interface} for more concrete illustrations with $s$ polarization.
\subsubsection{Hyperbolic cotangent}\label{subsec:coth}
In applying the fluctuation-dissipation theorem, we identified the autocorrelation with symmetrized quantum expectation: $\braket{JJ^*}\mapsto\braket{\hat{J}\hat{J}^\dag+\hat{J}^\dag\hat{J}}/2$, so that it is $\coth(\beta\hbar\omega/2)$ (or $2N(\omega)+1$ in terms of Bose function) that appears in \eqref{eq:JJ}--\eqref{eq:KK}. However, there are other variants of quantum expectation that one can choose. For example, the ``lesser'' $\braket{\hat{J}^\dag\hat{J}}$ (giving $2N$) and the ``greater'' $\braket{\hat{J}\hat{J}^\dag}$ (giving $2(N+1)$). Once B\"{u}ttiker symmetry is established, ultimately they all amount to the same term $2N$ because temperature-independent terms will be cancelled. Nevertheless, the symmetrized version is still preferred because it is the only odd function: $2N(\omega)+1=-[2N(-\omega)+1]$. This property is crucial for the validity of Def.~\eqref{eq:Poynting}.  
\subsubsection{Vacuum bulk}
When calculating the conductance, say between bulk bath 1 and interface $D$, $G_{1D}$, we started off with bulk 1 not being vacuum. Then, to check B\"{u}ttiker symmetry, we proceed to evaluate $G_{D1}$. But in this calculation, no assumption is needed for bulk 1: it is described by dielectric function $\epsilon_1$ which may very well be unity. Since these two are identical: $G_{1D}=G_{D1}$, we conclude that there is no harm in taking $\epsilon_1=1$ in the conductances $G_{1\alpha}$, provided that we take its temperature to be zero: $T_1=0$, so that the corresponding Bose function is zero: $N_1=0$, i.e. bath 1 only absorbs energy but does not radiate.
\subsubsection{Suspended sheets}\label{justif}
The radiative transfer between two suspended two-dimensional materials is studied in several works \cite{Ilic,Jiebin}. There, the transmission function corresponds to the $G_{0D}$ conductance in our work. Also, a Landauer-B\"{u}ttiker perspective provides alternate justification as regards the use of only evanescent mode in the calculations: one should picture having two vacuum bulk baths: $\epsilon_1=\epsilon_3=1$, to which energy carried by propagating mode flows. Indeed, thanks to the terms $\RE(\gamma)$ and $\RE(\gamma^*\epsilon)$ for $s$ and $p$ polarizations respectively, the conductances $G_{01},G_{03}$ (see Eqs.~\eqref{eq:conds} and \eqref{eq:condp}) are non-zero only for propagating mode ($|\qperp|>\omega/c$). Hence, for suspended sheets, the scattering center blocks small-wavelength ($|\qperp|<\omega/c$) channels, and only allows large-wavelengths to pass from the sheets to the bulks.
\subsubsection{Energy in vacuum gap}
For two objects separated by a vacuum gap, the radiative heat transfer between them is given by the Poynting vector at a point in the gap, $\mathcal{S}_C(d/2)$, say. To calculate this from a Landauer-B\"{u}ttiker perspective, we need to add the energies in links $1,0$ or $D,3$. More precisely, keeping the convention that energy entering bath be positive, we have:
\begin{equation}
\mathcal{S}_C\left(d/2\right)=-(\mathcal{S}_1+\mathcal{S}_0)=\mathcal{S}_3+\mathcal{S}_D,
\end{equation}
where the energy in each link $\mathcal{S}_\alpha$ can be evaluated using Eq. \eqref{eq:rht}.

\subsubsection{Uncorrelated distant noises}
The expression of total radiative heat flow in link $\alpha$, $\mathcal{S}_\alpha$ as given in Eq.~\eqref{eq:rht} suggests that one can add the energy contribution from each source separately. Indeed, we assume that sources from different systems are uncorrelated. Using bulk 1 and interface 0 for example, we have $\braket{\bm{J}^{(1)}\bm{K}^{(0)}}=\braket{\bm{J}^{(1)}}\braket{\bm{K}^{(0)}}$. Since the currents originate from thermal motions, their first moment is taken to be zero in average, hence $\braket{\bm{J}^{(1)}\bm{K}^{(0)}}=0$. This allows us to add energies from different sources independently.

\subsection{Example}
The four-terminal point of view also allows an easy bookkeeping of the energy dissipated in (or emitted by) the surfaces, simply by applying Kirchhoff's circuital law at nodes $0$ and $D$. We consider the radiative heat transfer between a gold slab covered by graphene, both kept at temperature $T_L$, with another suspended sheet of graphene at $T_R$:
\begin{figure}[H]
	\centering
	\includegraphics[scale=1]{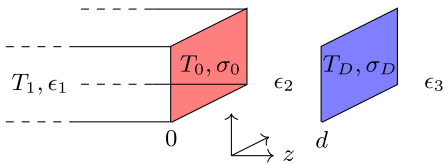}
	\caption{Radiative heat transfer between a gold slab covered by graphene kept at temperature $T_1=T_0=T_L$, and another suspended ($\epsilon_3=1$) graphene sheet at $T_D=T_R$, separated by a vacuum gap ($\epsilon_2=1$) at a distance $d$ apart. For mathematical convenience, the planar cross-sections are infinitely large.}
\end{figure}
For the dielectric function of gold, we use Drude model \cite{Ashcroft}: $\epsilon_1=1-ne^2/(m\varepsilon_0\omega[\omega+\ii/\tau])$, where $n$ is the carrier density, $\tau$ is the relaxation time, $m$ is the electron mass and $e$ is the elementary electron charge. As for the conductivities, we take $\sigma_0=\sigma_D=4\kappa E_F/(\pi\hbar[\Gamma-\ii\omega])$ for both graphene sheets \cite{graphene}, where $E_F$ is the Fermi energy, $\Gamma$ is a fitting parameter, and $\kappa=e^2/(4\hbar)$. The heat dissipated in the suspended graphene is given by:
\begin{equation}
\begin{split}
\mathcal{S}_D = \int_0^{\infty}&\frac{\mathrm{d}\omega}{2\pi}\hbar\omega
\int_{\mathbb{R}^2}\frac{\mathrm{d}^2\qperp}{(2\pi)^2}\Big[G_{1D}\left(N_1-N_D\right)\\&+G_{0D}(N_0-N_D)+G_{3D}(N_3-N_D)\Big],\label{eq:coatbulk}
\end{split}
\end{equation}
where the total conductance $G_{\alpha\delta}=G^s_{\alpha\delta}+G^p_{\alpha\delta}$ is given by the sum of the two polarizations (see Appendix \ref{app:Conductance} for explicit expressions). Since the graphene-gold system is kept at one single temperature $T_1=T_0=T_L$, we have $N_1=N_0$. On the other hand, the right graphene is suspended, meaning bulk bath 3 is a vacuum at zero temperature: $T_3=0$, leading to $N_3=0$. We split Eq. \eqref{eq:coatbulk} into $\mathcal{S}_D=\mathcal{S}_{1\to D}+\mathcal{S}_{0\to D}+\mathcal{S}_{3\to D}$ and plot the three contributions to $\mathcal{S}_D$ as function of gap separation $d$:
\begin{figure}[H]
	\includegraphics[width=1.0\columnwidth]{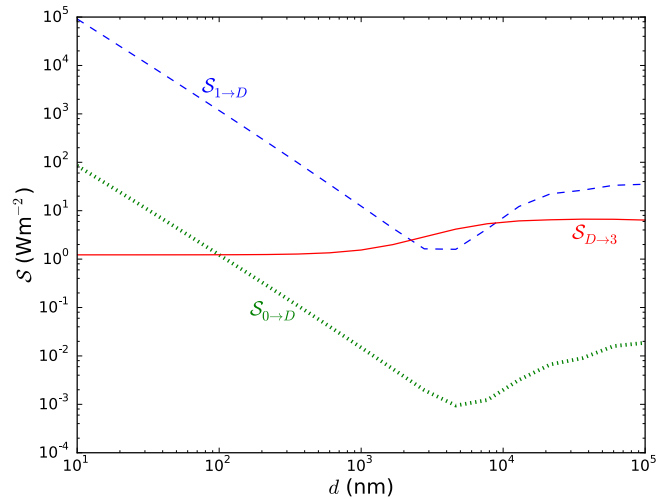}
	\caption{Radiative heat exchange with interface $D$ as function of gap separation $d$. Dashed line: from bulk 1. Dotted line: from interface 0. Solid line: to vacuum bulk 3 ($\mathcal{S}_{3\to D}<0$, graph shown is its magnitude $\mathcal{S}_{D\to 3}$). Parameters used are $T_L=373$ K, $T_R=273$ K, $n=5.9\times10^{22}$ cm$^{-3}$, $\tau=2.1\times10^{-14}$ s, $E_F=0.3$ eV, $\hbar\Gamma=3.7$ meV.}
\end{figure}
From the above, we identify each contribution to the radiative heat received or emitted by the suspended sheet $D$. First, the graphene-gold system, being at a higher temperature ($T_1=T_0>T_D$), always radiates to sheet $D$. Second, the presence of a vacuum bath ($T_3=0$) leads to energy loss from sheet $D$. In the near-field regime ($d<10^2$ nm say, with the parameters chosen), the energy absorptions from bulk 1 and interface 0 dominate the emission to vacuum bulk 3. Hence, for small gap separations, one can safely ignore the loss to vacuum bulk 3, which is nothing but the propagating mode (see Sec.~\ref{justif}). As the gap separation $d$ increases, the dashed and solid lines approach each other. This is where the loss to sink becomes comparable to the heat from source. In particular, when solid line is above dashed line, the suspended sheet $D$ is effectively radiating heat away to the vacuum. Similar analysis could be performed to estimate the gap separation above which one must take into account energy loss due to propagating mode. Furthermore, the radiative heat due to the graphene coat (dotted line) has a similar profile as the bulk gold (dashed line), albeit several orders smaller in magnitude. One possible reason that surfaces do not contribute as much in radiative transfer compared with bulk solids is the difference of dimensionality. Finally, for dashed and dotted lines, the linear part has a slope of minus two, in accordance with the $d^{-2}$ law for near-field radiative heat transfer \cite{Chen}.
\section{\label{sec:conclusion}CONCLUSION}
To summarize, we presented a Landauer-B\"{u}ttiker perspective for the radiative heat transfer in layered media with conductive interfaces. The bulk media and interfaces are regarded as bosonic baths, coupled perfectly to a scattering center (which constitutes of the interfaces themselves and a vacuum gap). The case of suspended sheets and graphene-coated media are both encompassed as special cases of the model. We provided explicit expressions for a conductance matrix, so that the energy exchange between each subsystem can now be written with ease. The parallels drawn also show that Rytov's fluctuational electrodynamics (with spatial isotropy and homogeneity) is ballistic \cite{Datta}, so that one has to incorporate spatial dispersion \cite{dispersion} or break local equilibrium hypothesis \cite{WangNF} to give way to novel phenomena.

\begin{acknowledgements}
The author thanks Jiebin Peng and Yi Wei Ho for fruitful discussions.
\end{acknowledgements}

\appendix

\section{MODIFIED FRESNEL COEFFICIENTS}
Here we provide some useful properties of the modified Fresnel coefficients \eqref{eq:mFresnels}---\eqref{eq:mFresnelp}. First, the following rules apply (hold also for the usual coefficients):
\begin{align}
\mt^s_{12}-\mr^s_{12}&=1, \\
\frac{\gamma_2\sqrt{\epsilon_1}}{\gamma_1\sqrt{\epsilon_2}}\mt^p_{12}+\mr^p_{12}&=1, \\
{\gamma_2}\mt^{s/p}_{12}&={\gamma_1}\mt^{s/p}_{21}.
\end{align}
Next, the properties below are different from the ones without conductive interfaces:
\begin{align}
\mr^s_{12}+\mr^s_{21} &= -\frac{2\mu_0\omega\sigma_0}{\gamma_1+\gamma_2+\mu_0\omega\sigma_0}, \\
\mr^p_{12}+\mr^p_{21} &= \frac{2\gamma_1\gamma_2\frac{\sigma_0}{\varepsilon_0\omega}}{\denompo},\\
\mt^s_{12}\mt^s_{21}-\mr^s_{12}\mr^s_{21} &= 1 + (\mr^s_{12}+\mr^s_{21}),\\
\mt^p_{12}\mt^p_{21}-\mr^p_{12}\mr^p_{21} &= 1 - (\mr^p_{12}+\mr^p_{21}).
\end{align}

\section{BULK}\label{app:Bulk}
Here we list the electric fields in all three regions due to fluctuating current in bulk bath 1. From these we compute the energies in each link, leading to explicit expressions for the conductance matrix. One could then show B\"{u}ttiker symmetry: $G_{\alpha\delta}=G_{\delta\alpha}$. We briefly outline the calculations for $s$ polarization.
Our goal is to evaluate the first column: $G^s_{10},G^s_{1D},G^s_{13}$, of the conductance matrix. To this end, we first switch off all but bath 1. Then, we calculate the electromagnetic fields in region 1, 2, 3. From these we compute the Poynting vector, and apply energy balance at nodes 0 and $D$ for the energy exchange with the interfaces. 
\subsection{Field}
\begin{figure}[H]
	\centering
	\includegraphics[scale=1]{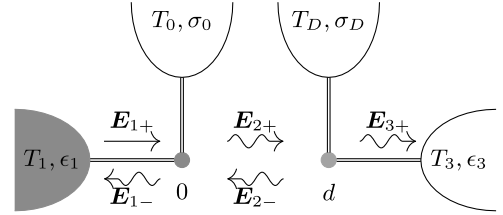}
	\caption{Scattered electric fields (wiggly arrows) due to incident field (straight arrow) from bulk 1 (shaded bath).}\label{fig:bulk}
\end{figure}

We begin by calculating the electromagnetic fields as illustrated in Fig.~\ref{fig:bulk}.
In above, the fields at $z<0$ ($z>d$) are evaluated at $z=0^-$ ($z=d^+$), whereas the middle ones can be taken at any point $z\in(0,d)$ because region 2 is vacuum. We have:
\begin{align}
\begin{split}
E_{s_{1-}}|_{z=0^-} &= \left[\mr^s_{12}+\frac{\mt^s_{12}\mt^s_{21}\mr^s_{23}\ee^{\ii 2\gamma_2 d}}{1-\mr^s_{23}\mr^s_{21}\ee^{\ii 2\gamma_2 d}}\right] E_{s_{1+}}\\&=\left[\frac{(1+\mr^s_{23}\expp)(1+\mr^s_{12})}{\Ds}-1\right]E_{s_{1+}},
\end{split} \\
E_{s_{2+}}|_{0<z<d} &= \frac{\mt^s_{12}\ee^{\ii\gamma_2z}}{1-\mr^s_{23}\mr^s_{21}\ee^{\ii 2\gamma_2 d}}E_{s_{1+}}, \\
E_{s_{2-}}|_{0<z<d} &= \frac{\mt^s_{12}\mr^s_{23}\ee^{\ii \gamma_2 (2d-z)}}{1-\mr^s_{23}\mr^s_{21}\ee^{\ii 2\gamma_2 d}} E_{s_{1+}}, \\
E_{s_{3+}}|_{z=d^+} &= \frac{\mt^s_{12}\mt^s_{23}\ee^{\ii\gamma_2d}}{1-\mr^s_{23}\mr^s_{21}\ee^{\ii 2\gamma_2 d}}E_{s_{1+}},
\end{align}
where $E_{s_{1+}}=-{\mu_0\omega}J_{s_1}/({2\gamma_1})$, as can be seen from the Green's dyadic \eqref{eq:GE}.

\subsection{Energy}
By definition (Eq. \eqref{eq:rht}), when one works in Fourier domain $(\qperp,\omega)$, the factor relating mean oscillator energy $\hbar\omega N(\omega)$ to the electromagnetic energy $S(\qperp,\omega)$ is identified as the conductance. Hence, we evaluate $2\braket{\RE(\bm{E}\times\bm{H}^*)}$, remove from it $\hbar\omega\coth(\beta\hbar\omega/2)$ and multiply by two (see Sec.~\ref{subsec:coth}) to obtain the conductance. To calculate the spectral Poynting vector using \eqref{eq:JJ}, one needs to contract two dyadics using $\delta_{ij}$, apply convolution theorem, simplify double integrals with delta functions, and integrate along $z$ axis. The conclusion is, one can effectively work with a single Fourier component $(\qperp,\omega)$, and apply the following replacements:
\begin{align}
	\begin{split}
|J_s|^2&\mapsto \frac{\hbar}{\mu_0}\RE(\gamma)\coth\left[\frac{\beta\hbar\omega}{2}\right],\\
|J_p|^2&\mapsto \frac{\hbar}{\mu_0}\RE(\gamma)\coth\left[\frac{\beta\hbar\omega}{2}\right]|\hatp|^2.\label{eq:replJ}
\end{split}
\end{align}By planar symmetry, only the $z$ component of the Poynting vector contributes to radiative heat transfer. In region $i$, if the field travels along a single direction, taking $z$ component amounts to appending $\RE(\gamma_i)$ and ${|\epsilon_i|}^{-1}{\RE(\gamma_i^*\epsilon_i)}$ for $s$ and $p$ polarization respectively. Below we denote by $D_s$ the Fabry-P\'{e}rot denominator for $s$ polarization: $D_s=\Ds$, describing multiple reflections between two planes. The notation $S^s_{\alpha\to\delta}$ means we consider $s$ polarization of the energy from system $\alpha$ to $\delta$, where $\alpha,\delta\in\{1,0,D,3,C\}$ are indices referring to bulk, interface or the center.
\subsubsection{$S^s_{1\to 1}$}
In link 1, the incident field $E_{s_{1+}}$ (obtained by integrating over bath 1) interferes with the reflected field $E_{s_{1-}}$. Thus, the expression of energy in link 1, due to fluctuating source in bulk 1, which we call $S^s_{1\to 1}$, is slightly more complicated: 
\begin{align}
	\begin{split} 
2&\braket{\RE(\bm{E}_1\times\bm{H}_1^*)_z}=\frac{\mu_0\omega}{2|\gamma_1|^2}|J_{s_1}|^2 \\ &\times \left[\RE(\gamma_1)(1-|\mr^s_{13}|^2)+2\IM(\mr^s_{13})\IM(\gamma_1)\right],\label{s:1}
	\end{split}
\end{align}
where the effective modified reflection coefficient from bulk 1 to bulk 3 is given by:
\begin{equation}
\mr^s_{13} = \mr^s_{12}+\frac{\mt^s_{12}\mt^s_{21}\mr^s_{23}\ee^{\ii 2\gamma_2 d}}{1-\mr^s_{23}\mr^s_{21}\ee^{\ii 2\gamma_2 d}}.
\end{equation}
With some elementary calculations, from \eqref{s:1} we obtain:
\begin{widetext} 
\begin{align}
\begin{split} 
&2\braket{\RE(\bm{E}_1\times\bm{H}_1^*)_z}\\&=\frac{\mu_0\omega}{2|\gamma_1|^2}|J_{s_1}|^2\left\{ 2\RE\left[\gamma_1^*\frac{1+\mr^s_{23}\expp}{D_s}\mt^s_{12}\right]-\RE(\gamma_1)\left|\frac{1+\mr^s_{23}\expp}{D_s}\mt^s_{12}\right|^2\right\}\\
&=\frac{2\mu_0\omega|J_{s_1}|^2}{|D_s(\gamma_1+\gamma_2+\mu_0\omega\sigma_0)|^2}\Bigg\{\RE(\gamma_2^*)(1-|\mr^s_{23}\expp|^2)+2\IM(\gamma_2)\IM(\mr^s_{23}\expp)+\mu_0\omega\RE(\sigma_0^*)|1+\mr^s_{23}\expp|^2\Bigg\}.
\end{split} 
\end{align}
\end{widetext}
Since $\epsilon_2=1$, $\gamma_2$ is either real or purely imaginary. This suggests us to discuss separately propagating ($|\qperp|<\omega/c$ hence $\gamma_2=|\gamma_2|$) and evanescent mode ($|\qperp|>\omega/c$ hence $\gamma_2=\ii|\gamma_2|$). It turns out that there is a way to write both cases:
\begin{equation}
\begin{split} 
&\frac{4|\gamma_2|^2|\expp|[\RE(\gamma_3)+\mu_0\omega\RE(\sigma_D)]}{|\denomsd|^2}\\&=
\begin{cases}
\RE(\gamma_2)(1-|\mr^s_{23}\expp|^2), &\textrm{if }|\qperp|<\frac{\omega}{c}, \\
2\IM(\gamma_2)\IM(\mr^s_{23}\expp), &\textrm{if }|\qperp|>\frac{\omega}{c}.
\end{cases}
\end{split} 
\end{equation}
Therefore, we obtain the energy due to bulk 1, flowing in the link between bath 1 and the center:
\begin{equation}
\begin{split} 
&S^s_{1\to 1}=\frac{2\mu_0\omega|J_{s_1}|^2}{|D_s(\gamma_1+\gamma_2+\mu_0\omega\sigma_0)|^2}\\ &\times \Bigg\{\frac{4|\gamma_2|^2|\expp|[\RE(\gamma_3)+\mu_0\omega\RE(\sigma_D)]}{|\denomsd|^2}\\ &+\mu_0\omega\RE(\sigma_0^*)|1+\mr^s_{23}\expp|^2\Bigg\}.
\end{split}
\end{equation}
\subsubsection{$S^s_{1\to C}$}
This term represents the energy flowing in the vacuum gap due to bulk 1:
\begin{align}
\begin{split}
&2\braket{\RE(\bm{E}_2\times\bm{H}_2^*)_z}\\&=\frac{2\mu_0\omega|J_{s_1}|^2}{|D_s(\denomso)|^2}\Bigg\{\RE(\gamma_2^*)(1-|\mr^s_{23}\ee^{\ii\gamma_2(2d-z)}|^2)\\&+2\IM(\gamma_2)\IM\left[\mr^s_{23}\ee^{\ii2\gamma_2 d}\ee^{-2\ii\RE(\gamma_2)z}\right]\Bigg\}.
\end{split}
\end{align}
After some calculations the above becomes:
\begin{equation} 
\frac{2\mu_0\omega|J_{s_1}|^2}{|D_s(\denomso)|^2}\frac{4|\gamma_2|^2|\expp|[\RE(\gamma_3)+\mu_0\omega\RE(\sigma_D)]}{|\denomsd|^2}.
\end{equation} 
\subsubsection{$S^s_{1\to 3}$}
A source in bulk 1 generates only a forward-going field, without back-scattered term in bulk 3. Thus the energy flowing in link 3 is easy to calculate:
\begin{equation}
\begin{split} 
&2\braket{\RE(\bm{E}_3\times\bm{H}^*_3)_z} \\&= \frac{8\mu_0\omega|\gamma_2|^2|\expp||J_{s_1}|^2\RE(\gamma_3)}{|D_s|^2|(\denomso)(\denomsd)|^2}.
\end{split}
\end{equation}
We can now obtain the conductance element $G^s_{10}$, by applying the fluctuation-dissipation theorem to $|J_{s_1}|^2$, removing $\hbar\omega\coth\left[{\beta_1\hbar\omega}/{2}\right]$, and bringing in a factor of 2:
\begin{equation}
G^s_{13} = \frac{16|\gamma_2|^2|\expp|\RE(\gamma_1)\RE(\gamma_3)}{|D_s(\denomso)(\denomsd)|^2}.
\end{equation}
This expression is unaffected under the exchange $1\leftrightarrow3$ and $0\leftrightarrow D$, thus we deduce immediately the first B\"{u}ttiker symmetry: $G^s_{13}=G^s_{31}$.
\subsubsection{$S^s_{1\to 0}$}
We can now calculate the energy flowing to interface 0 due to a source in bulk 1, $S^s_{1\to 0}$ by applying Kirchhoff circuital law at node 0:
\begin{equation}
\begin{split}
S^s_{1\to 0}&=S^s_{1\to 1}-S^s_{1\to C} \\
&=\frac{2\mu_0\omega|J_{s_1}|^2\mu_0\omega\RE(\sigma_0^*)|1+\mr^s_{23}\expp|^2}{|D_s(\gamma_1+\gamma_2+\mu_0\omega\sigma_0)|^2},
\end{split}
\end{equation}
which gives the conductance:
\begin{equation}\label{G10}
G^s_{10}=\frac{4|1+\mr^s_{23}\expp|^2\RE(\gamma_1)\mu_0\omega\RE(\sigma_0)}{|D_s(\gamma_1+\gamma_2+\mu_0\omega\sigma_0)|^2}.
\end{equation}
\subsubsection{$S^s_{1\to D}$}
As before, to calculate the energy exchange with interface $D$ due to a source in bulk 1, one applies Kirchhoff's law at node $D$:
\begin{equation}
\begin{split}
S^s_{1\to D} &= S^s_{1\to C}-S^s_{1\to 3} \\
&=\frac{8\mu_0\omega|\gamma_2|^2|\expp||J_{s_1}|^2\mu_0\omega\RE(\sigma_D)}{|D_s|^2|(\denomso)(\denomsd)|^2},
\end{split} \label{eq:S1D}
\end{equation}
whence the conductance:
\begin{equation}\label{G1d}
G^s_{1D}=\frac{16|\gamma_2|^2|\expp|\RE(\gamma_1)\mu_0\omega\RE(\sigma_D)}{|D_s(\denomso)(\denomsd)|^2}.
\end{equation}

\section{INTERFACE}\label{app:Interface}
We consider here the electromagnetic field and energy due to fluctuating current on interface 0. This means we are now interested in the second column: $G^s_{01},G^s_{0D},G^s_{03}$ of the conductance matrix in Eq.~\eqref{eq:mat}.  
\subsection{Field}
\begin{figure}[H]
	\centering
	\includegraphics[scale=1]{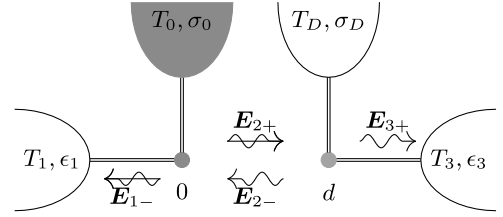}
	\caption{Scattered electric fields (wiggly arrows) due to fluctuation-induced fields (straight arrow) from interface 0 (shaded bath).}\label{fig:interface}
\end{figure}

To calculate Poynting vector, we first solve a scattering problem of electric fields as shown in Fig.~\ref{fig:interface}.
Unlike the previous case where we can regard the field from bulk 1 as distinctively incident, here the scattering is intertwined with emission. Thus we need to express the fields in terms of the fluctuation $\bm{K}_0$:
\begin{align}
\begin{split}
	E_{s_{1-}}|_{z=0^-}&=-\mu_0\omega\left[1+\frac{\mr^s_{23}\mt_{21}^s\expp}{\Ds}\right]\frac{\mt^s_{12}}{2\gamma_1}K_{s_0} \\
	&=-\mu_0\omega\left[\frac{1+\mr^s_{23}\expp}{\Ds}\right]\frac{\mt^s_{12}}{2\gamma_1}K_{s_0},
\end{split}\\
	E_{s_{2+}}|_{0<z<d}&=-\mu_0\omega \left[\frac{\ee^{\ii\gamma_2z}}{\Ds}\right]\frac{\mt^s_{12}}{2\gamma_1}K_{s_0}, \\
	E_{s_{2-}}|_{0<z<d}&=-\mu_0\omega\left[\frac{\mr^s_{23}\ee^{\ii\gamma_2(2d-z)}}{\Ds}\right]\frac{\mt^s_{12}}{2\gamma_1}K_{s_0}, \\
	E_{s_{3+}}|_{z=d^+}&=-\mu_0\omega\left[\frac{\mt^s_{23}\ee^{\ii\gamma_2d}}{\Ds}\right]\frac{\mt^s_{12}}{2\gamma_1}K_{s_0}.
\end{align}

\subsection{Energy}
From electric fields, we calculate the radiative heat using the expression $2\RE\braket{(\bm{E}\times\bm{H}^*)_z}$. One essential step is still to apply the fluctuation-dissipation theorem \eqref{eq:KK}. Since the fluctuating surface current stays strictly in plane and we demand planar isotropy, the replacements analogous to \eqref{eq:replJ} are given by:
\begin{equation}
|K_s|^2,|K_{q_\perp}|^2 \mapsto \hbar\omega\RE(\sigma)\coth\left[\frac{\beta\hbar\omega}{2}\right].\label{eq:replK}
\end{equation}
\subsubsection{$S^s_{0\to 1}$}
We begin with the energy originated from interface 0, flowing in the link to bath 1:
\begin{align}
\begin{split}
&2\braket{\RE(\bm{E}_1\times\bm{H}_1^*)_z}\\&=\frac{2\mu_0\omega}{|(\denomso)D_s|^2}|1+\mr^s_{23}\expp|^2|K_{s_0}|^2\RE(\gamma_1).
\end{split}
\end{align}
From this we extract the conductance element $G^s_{01}$:
\begin{equation}
G^s_{01}=\frac{4|1+\mr^s_{23}\expp|^2\RE(\gamma_1)\mu_0\omega\RE(\sigma_0)}{|(\denomso)(\Ds)|^2},
\end{equation}
which is identical to \eqref{G10}. Thus we verify the second B\"{u}ttiker symmetry: $G^s_{01}=G^s_{10}$. We considered bulk 1 and interface 0, but we could have started instead with bulk 3 and interface $D$. Hence, we see that $G^s_{D3}=G^s_{3D}$.
\subsubsection{$S^s_{0\to C}$}
We now calculate the energy from interface 0 to the center:
\begin{align}
\begin{split}
&2\braket{\RE(\bm{E}_2\times\bm{H}_2^*)_z}=\frac{2\mu_0\omega|K_{s_0}|^2}{|D_s(\denomso)|^2}\\ &\times\Bigg\{\RE(\gamma_2^*)(1-|\mr^s_{23}\ee^{\ii\gamma_2(2d-z)}|^2)\\&+2\IM(\gamma_2)\IM\left[\mr^s_{23}\ee^{\ii2\gamma_2 d}\ee^{-2\ii\RE(\gamma_2)z}\right]\Bigg\}.
\end{split}
\end{align}
After some manipulations, the above reduces to:
\begin{equation} 
\frac{8\mu_0\omega|K_{s_0}|^2|\gamma_2|^2|\expp|[\RE(\gamma_3)+\mu_0\omega\RE(\sigma_D)]}{|D_s(\denomso)(\denomsd)|^2}.
\end{equation}

\subsubsection{$S^s_{0\to 3}$}
To apply Kirchhoff's law at node $D$, we calculate the energy from interface 0 to bath 3:
\begin{equation}
\begin{split}
S^s_{0\to 3}&=\frac{8\mu_0\omega|\gamma_2|^2|\expp|\RE(\gamma_3)|K_{s_0}|^2}{|D_s(\denomso)(\denomsd)|^2}. \label{eq:S03}
\end{split}
\end{equation}
Earlier, we calculated the energy from bulk 1 to interface $D$, c.f. Eq.~\eqref{eq:S1D}. Hence, we need instead $S^s_{D\to 1}$ to check B\"{u}ttiker symmetry. This can be achieved by a simple replacement $1\leftrightarrow 3$ in \eqref{eq:S03}. Therefore, only the autocorrelation term $|K_{s_0}|^2$, and the $z$ component of wave-vector, $\gamma_3$, are affected. After the index replacement we find the conductance matrix element:
\begin{equation}
G^s_{D1}= \frac{16|\gamma_2|^2|\expp|\RE(\gamma_1)\mu_0\omega\RE(\sigma_D)}{|D_s(\denomso)(\denomsd)|^2}.
\end{equation}
Referring back to \eqref{G1d}, we checked the fourth symmetry: $G^s_{D1}=G^s_{1D}$. The case of interface $0$ and bulk 3 is exactly the same with index replacements $D\leftrightarrow 0$ and $1\leftrightarrow 3$. Thus we conclude that $G^s_{03}=G^s_{30}$.

\subsubsection{$S^s_{0\to D}$}
This term represents the energy from interface 0 to interface $D$, and is calculated by energy balance at node $D$:
\begin{align}
\begin{split}
S^s_{0\to D} &= S^s_{0\to C} - S^s_{0\to 3} \\
&=\frac{8(\mu_0\omega)^2|\gamma_2|^2|\expp|\RE(\sigma_D)|K_{s_0}|^2}{|D_s(\denomso)(\denomsd)|^2}.
\end{split}
\end{align}
Substituting the autocorrelation function by \eqref{eq:replK}, we find the conductance:
\begin{equation}
G^s_{0D} = \frac{16|\gamma_2|^2|\expp|[\mu_0\omega\RE(\sigma_D)][\mu_0\omega\RE(\sigma_0)]}{|D_s(\denomso)(\denomsd)|^2}.
\end{equation}
Clearly, the expression above is unchanged under the permutations $1\leftrightarrow 3$ and $0\leftrightarrow D$, thus the last B\"{u}ttiker symmetry: $G^s_{0D}=G^s_{D0}$ is verified.

\section{CONDUCTANCE MATRIX}\label{app:Conductance}
Here we collect and list the lower-triangle part of the conductance matrix in \eqref{eq:mat}. Having established B\"{u}ttiker symmetry, the upper-triangle part is thus identical. Finally, the diagonal elements are given by $G_{\alpha\alpha}=-\sum_{\delta\neq\alpha} G_{\delta\alpha}$ thanks to energy balance.
\subsection{$s$ polarization}
The conductance elements for $s$ polarization are given by:
\begin{widetext}
	\begin{align}
			G^s_{10} &=\frac{4|1+\mr^s_{23}\expp|^2}{|D_s(\gamma_1+\gamma_2+\mu_0\omega\sigma_0)|^2}\RE(\gamma_1)\mu_0\omega\RE(\sigma_0), \\
			G^s_{1D} &=\frac{16|\gamma_2|^2|\expp|}{|D_s(\denomso)(\denomsd)|^2}\RE(\gamma_1)\mu_0\omega\RE(\sigma_D), \\
			G^s_{13} &= \frac{16|\gamma_2|^2|\expp|}{|D_s(\denomso)(\denomsd)|^2}\RE(\gamma_1)\RE(\gamma_3), \\
			G^s_{0D} &= \frac{16|\gamma_2|^2|\expp|}{|D_s(\denomso)(\denomsd)|^2}[\mu_0\omega\RE(\sigma_0)][\mu_0\omega\RE(\sigma_D)], \\
			G^s_{03} &= \frac{16|\gamma_2|^2|\expp|}{|D_s(\denomso)(\denomsd)|^2}\mu_0\omega\RE(\sigma_0)\RE(\gamma_3),\label{eq:conds}\\
			G^s_{D3} &= \frac{4|1+\mr^s_{21}\expp|^2}{|D_s(\denomsd)|^2}\mu_0\omega\RE(\sigma_D)\RE(\gamma_3).
	\end{align}
\end{widetext}
	\subsection{$p$ polarization}
	The calculation of $p$ polarized conductance matrix is similar so we shall only provide the expressions. As before, we denote by $D_p=\Dp$ the Fabry-P\'{e}rot denominator for $p$ polarization. We have:
	\begin{widetext}
	\begin{align}
		G^p_{10} &= \frac{4|1-\mr^p_{23}\expp|^2}{|D_p(\denompo)|^2}\RE(\gamma_1^*\epsilon_1)|\gamma_2|^2\frac{\RE(\sigma_0)}{\varepsilon_0\omega}, \\
		G^p_{1D} &= \frac{16|\gamma_2|^2|\expp|}{|D_p(\denompo)(\denompd)|^2}\RE(\gamma_1^*\epsilon_1)|\gamma_3|^2\frac{\RE(\sigma_D)}{\varepsilon_0\omega}, \\
		G^p_{13} &= \frac{16|\gamma_2|^2|\expp|}{|D_p(\denompo)(\denompd)|^2}\RE(\gamma_1^*\epsilon_1)\RE(\gamma_3^*\epsilon_3),\\
		G^p_{0D} &= \frac{16|\gamma_2|^2|\expp|}{|D_p(\denompo)(\denompd)|^2}|\gamma_1|^2\frac{\RE(\sigma_0)}{\varepsilon_0\omega}|\gamma_3|^2\frac{\RE(\sigma_D)}{\varepsilon_0\omega},\\
		G^p_{03} &= \frac{16|\gamma_2|^2|\expp|}{|D_p(\denompo)(\denompd)|^2}|\gamma_1|^2\frac{\RE(\sigma_0)}{\varepsilon_0\omega}\RE(\gamma_3^*\epsilon_3),\label{eq:condp}\\
		G^p_{D3} &= \frac{4|1-\mr^p_{21}\expp|^2}{|D_p(\denompd)|^2}|\gamma_2|^2\frac{\RE(\sigma_D)}{\varepsilon_0\omega}\RE(\gamma_3^*\epsilon_3).
	\end{align}
\end{widetext}
One can then use the above to write the energy exchange, due to either polarization, between any two subsystems, in a similar way to Eq. \eqref{eq:rht}.

\end{document}